\documentclass[fleqn,usenatbib]{mnras}

\usepackage{newtxtext,newtxmath}

\usepackage[T1]{fontenc}
\usepackage{ae,aecompl}

\usepackage{graphicx}	
\usepackage{amsmath}	
\usepackage{pifont}
\usepackage[table]{xcolor} 
\usepackage[normalem]{ulem} 
\usepackage{CJKutf8}
\usepackage{subcaption}

\definecolor{japviolet}{rgb}{0.53, 0., 0.69}

\definecolor{ejcol}{rgb}{0.8,0,1}

\definecolor{jhcol}{rgb}{0,0.6,0}

\title[Machine-guided Simulation Calibration]{Machine-guided Exploration and Calibration of Astrophysical Simulations}

\author[BK. Oh et al.]{%
Boon Kiat Oh,$^{1}$\thanks{Corresponding authors: \href{mailto:bkoh1986@snu.ac.kr}{bkoh1986@snu.ac.kr}}
Hongjun An,$^{1, 2}$
Eun-jin Shin,$^{1}$
Ji-hoon Kim $^{1, 3}$\thanks{Corresponding authors: \href{mailto:me@jihoonkim.org}{me@jihoonkim.org}}
\newauthor and Sungwook E. Hong \begin{CJK*}{UTF8}{mj}(홍성욱)\end{CJK*}$^{4,5}$
\\
\\
$^{1}$Center for Theoretical Physics, Department of Physics and Astronomy, Seoul National University, Seoul 08826, Korea
\\
$^{2}$Department of Electrical and Computer Engineering, Seoul National University, Seoul 08826, Korea
\\
$^{3}$Seoul National University Astronomy Research Center, Seoul 08826, Korea
\\
$^{4}$Korea Astronomy and Space Science Institute, 776 Daedeok-daero, Yuseong-gu, Daejeon 34055, Korea
\\
$^{5}$Astronomy Campus, University of Science \& Technology, 776 Daedeok-daero, Yuseong-gu, Daejeon 34055, Korea
}

\date{Accepted XXX. Received YYY; in original form ZZZ}

\pubyear{2022}

\begin{document}
\label{firstpage}
\pagerange{\pageref{firstpage}--\pageref{lastpage}}
\maketitle

\begin{abstract}
We apply a novel method with machine learning to calibrate sub-grid models within numerical simulation codes to achieve convergence with observations and between different codes. It utilizes active learning and neural density estimators. The hyper parameters of the machine are calibrated with a well-defined projectile motion problem. Then, using a set of 22 cosmological zoom simulations, we tune the parameters of a popular star formation and feedback model within {\tt Enzo} to match simulations. The parameters that are adjusted include the star formation efficiency, coupling of thermal energy from stellar feedback, and volume into which the energy is deposited. This number translates to a factor of more than three improvements over manual calibration. Despite using fewer simulations, we obtain a better agreement to the observed baryon makeup of a Milky-Way (MW) sized halo. Switching to a different strategy, we improve the consistency of the recommended parameters from the machine. Given the success of the calibration, we then apply the technique to reconcile metal transport between grid-based and particle-based simulation codes using an isolated galaxy. It is an improvement over manual exploration while hinting at a less known relation between the diffusion coefficient and the metal mass in the halo region. The exploration and calibration of the parameters of the sub-grid models with a machine learning approach is concluded to be versatile and directly applicable to different problems.   
\end{abstract}

\begin{keywords}
galaxies:formation -- galaxies:evolution -- galaxies:haloes
\end{keywords}



\section{Introduction}

Dark matter and baryons are the drivers of structure formation and evolution in the universe. The interplay between the two components has been studied with highly accurate numerical simulations. As the computational capability advanced, there has been corresponding improvements in the resolution over the years \citep{1985ApJS...57..241E, 1999ApJ...524L..19M, 2008MNRAS.391.1685S, 2011ApJ...740..102K}. Despite this improvement, the resolution required to resolve baryonic processes such as star formation, feedback from stars, and black holes in a cosmological setting remains a challenge. Instead, subgrid models are used to represent these processes in simulations \citep{2003MNRAS.339..289S, 2010Natur.463..203G, 2013ApJ...770...25A, 2019MNRAS.484.2632S}. For the simulations to have predictive power, the parameters of the subgrid models has to be calibrated to reproduce observations \citep{2014PASJ...66...70O, 2015MNRAS.446..521S, 2020MNRAS.497.5203O}.

Feedback stems from the star formation and active galactic nuclei (AGN) activity \citep{2007MNRAS.380..877S, 2009MNRAS.398...53B, 2011MNRAS.414..195T}. It is an important component and can drive the relation between stellar mass and metallicity \citep{Rossi2007, Dave2011}. The outflow from the supernova (SN) feedback can also efficiently transports the metal-enriched materials away from the innermost region of the galaxy, regulating the metallicity \citep{Larson1974, Tremonti2004, Rossi2017}. Therefore, feedback plays a pivotal role in recreating realistic galaxies in simulations.

Each form of feedback has a different degree of impact depending on the mass of the halo. To implement them in numerical simulations depends on the various interpretation of these processes. Across different simulation codes, there is a variety of models for SN feedback alone \citep{2006ApJ...650..560C, 2008A&A...477...79D, 2012MNRAS.426..140D, 2018MNRAS.478..302S}. It can be designed as a thermal injection of energy or mechanical feedback \citep{2015MNRAS.451.2900K, 2018MNRAS.477.1578H}. The latter is found to be relatively resolution-independent and ensures convergence, provided that the resolution of the simulations are high enough.

Given the huge variation in implementation, there is a corresponding difference in the properties of galaxies in simulations \citep{2000ApJ...545..728T, 2003MNRAS.339..289S, 2005MNRAS.363.1299O, 2006MNRAS.373.1265O, 2010MNRAS.402.1536S}. This diversity is further compounded by the discrepancy in the adopted simulation codes, which can broadly be classified as grid-based or particle-based. Essentially, the difference lies in the way that gas is treated in the simulation, either deposited on a grid or modeled as a particle. 

The {\it AGORA} project was born to investigate both the convergence and divergence of the various simulation codes \citep{AGORA2014, AGORA2016}. One of the key findings of \citet{AGORA2016} points to the discrepancy in how metal is transported around the simulation domain by different simulation codes. In a grid-based code, metal enrichment of the hot diffuse gas can be found in the halo region due to stronger outflows and non-zero gas density present in the cells. However, a lack of gas in the halo region in particle-based codes, which is represented by the absence of particles leads to metal enrichment confined to the dense star-forming region. It restricts the metal enriched outflows to be caused by ram pressure. This difference can be alleviated by the inclusion of an explicit metal diffusion scheme, adding gas particles to the halo region and boosting feedback energy in particle-based codes \citep{2021ApJ...917...12S}. Therefore, feedback plays a pivotal role in simulations. Not only does it determine the baryon makeup within haloes, it provides the means to distribute the metals around the galaxy. Due to the nonlinear nature of how the properties of a simulated galaxy is shaped by the choice of feedback parameters, it is important to explore and understand the impact of the parameter space in these subgrid models. 

Amidst all the uncertainty introduced by the different simulation codes and subgrid routines, there exists a need to calibrate certain aspects of the simulation to observations \citep{2014PASJ...66...70O, 2015MNRAS.446..521S}. This step is necessary so that the simulation can serve as a testbed for non-calibrated features and possess predictive power. For instance, the `Evolution and Assembly of GaLaxies and their Environments' (EAGLE) simulation project is calibrated to reproduce several observed properties such as the $z=0.1$ galaxy stellar mass function (GSMF), the relation between the mass of galaxies and their central black holes and realistic galaxy sizes \citep{2015MNRAS.446..521S}. \citet{2020MNRAS.497.5203O} calibrated their suite of simulation to reproduce the baryon makeup of a Milky-Way (MW) sized galaxy at $z=0$, fueled by the observations from \citet{mcgaugh2009baryon}. These calibrations are costly and prohibit a complete exploration of parameter space. Hence, there is a need to streamline and focus the resources on regions of parameter space most likely to yield a desirable outcome. Given the nonlinear nature of the problem, it is difficult to identify the mentioned region, which motivates us to apply machine learning methods. 

Machine learning has been gaining a lot of attention as a useful tool that can be applied to a wide range of fields. It has yielded great results in different studies such as pattern recognition \citep{noh2015learning, sermanet2013pedestrian}, ill-posed inverse problem \citep{fessler2010model, yoon2018quantitative} and properties estimation \citep{papamakarios2019sequential,jo2019machine}. It has also been used to generate high resolution simulations from low resolution runs \citep{2021PNAS..11822038L, 2021MNRAS.507.1021N} and predict cosmological parameters from dark matter particles' positions and velocities in a simulation \citep{2018arXiv180804728M}. These achievements highlighted an improvement over conventional methods. However, machine learning methods are usually fueled by large amounts of data and the performance is enhanced by the availability of more data. This requirement demands a large amount of computation resources and is time-consuming, which is not feasible for the calibration problem we are interested in.

However, \citet{asling2019fast} proposed a density-estimation likelihood-free inference (DELFI) method, which used a small number of data points by utilizing active learning and neural density estimator. Active learning is one of the machine learning techniques that reduce the demand on the amount of required data \citep{settles.tr09}. This technique allows a more effective training through the interaction between the machine and data generation to determine which additional data is needed. This technique is particularly applicable in our case due to the prohibitively expensive nature of simulations.

In this paper, we propose a new and fast method for calibrating the sub-grid models of numerical simulations using machine learning approaches. We apply it to calibrate the baryon makeup of a MW-sized halo at $z=0$, identical to \citet{2020MNRAS.497.5203O}. We will compare the outcome in terms of the degree of agreement to observations and the amount of resources spent. Combining the results, we will discuss the ability of the model introduced by \citet{1992ApJ...399L.113C} and implemented in {\tt Enzo} \citep{2014ApJS..211...19B} to match observations. Also, we will apply the method to reconcile the metal transport between a grid-based and a particle-based code as described in \citet{2021ApJ...917...12S}. We will probe the versatility of the machine to be applied to different problems. Similar to before, we compare the degree of agreement and resources spent using machine learning methods to manual calibration. 

This paper is structured as follows. Section \ref{sec:methods} outlines the initial conditions of the simulations, simulation code, and setups to evolve the simulations. Also, we explain the methodology of the machine learning technique. In Section \ref{sec:results}, we calibrate the hyper parameters of our machine with a simplified problem: a projectile motion with air resistance. The optimized setup is then applied to the calibration of the star formation and feedback model and to resolve the metal transport discrepancy between grid-based and particle-based code. We will present how the new technique reduces the demand for computational resources and improves the agreement between the simulation and observations, and between different simulation codes. The limitations highlighted by the machine will also be discussed. Lastly, we summarise the paper with discussions and future works in Section \ref{sec:conclusions}.

\section{Methodology} \label{sec:methods}

\subsection{Simulations and their subgrid parameters}

\subsubsection{Feedback calibration in cosmological simulation} \label{sec:bk_setup}

This section provides a summary of the simulation setup which is identical to that presented in \citet{2020MNRAS.497.5203O} and \citet{2021arXiv210302234K}. We focus on a MW-sized halo in which the AGN feedback is relatively less significant \citep{2006MNRAS.370..645B, 2010ApJ...717..379B, 2014IAUS..303..354S}. We adopt a cosmology consistent with WMAP-9 \citep{2013ApJS..208...20B} and the parameters are $\Omega_m=0.285$, $\Omega_\Lambda=0.715$, $\Omega_b=0.0461$, $h = 0.695$ and $\sigma_8=0.828$ with the usual definitions. The initial conditions are generated with the MUlti-scale Initial Conditions ({\tt MUSIC}) for cosmological simulations \citep{2011MNRAS.415.2101H} and the zoom simulations are derived from the parent simulation with a volume of $L=100\,h^{-1}\,\mathrm{cMpc}$ with $256^3$ particles.

We evolve the simulations with an adaptive mesh-refinement (AMR) code \citep{2014ApJS..211...19B}, {\tt Enzo} and couple the {\tt ZEUS} \citep{1992ApJS...80..753S} hydro solver with an N-body adaptive particle-mesh gravity solver \citep{1985ApJS...57..241E}. The chemistry and cooling processes are handled by the {\tt Grackle} library \citep{2017MNRAS.466.2217S}. In particular, we employ the equilibrium cooling mode from {\tt Grackle}, which utilizes the tabulated cooling rates derived from the photoionization code CLOUDY \citep{2013RMxAA..49..137F} together with the heating from UV background radiation given by \citet{2012ApJ...746..125H}. 

We will focus on Setup 2 adopted in \citet{2020MNRAS.497.5203O}. The calibration is based on the modified star formation and feedback model described in \citet{1992ApJ...399L.113C}, looking primarily at $f_*$, $\epsilon$ and $r\_s$. These parameters can be broadly interpreted as star formation efficiency, a proxy for feedback energy budget and extent of feedback injection. We will define these parameters in greater details when they are mentioned in the star formation and feedback processes that follows. They will influence the baryon makeup in the halo at $z=0$. With three nested levels and five additional levels of AMR, the simulation has a dark matter particle mass resolution of $1.104\times10^7\,M_\odot$ and a maximum spatial resolution of 2.196 comoving kpc (ckpc).  

There are several conditions that deem a cloud of gas capable of forming a star:

\begin{enumerate}
	\item No further refinement within the cell
	\item Gas density greater than a threshold density: $\rho_\mathrm{gas} > \rho_\mathrm{threshold}$
	\item Convergent flow: $\boldsymbol{\nabla} \cdot {\bf v} < 0$
	\item Cooling time less than a dynamical time: $t_\mathrm{cool} < t_\mathrm{dyn}$
	\item Gas mass larger than the Jeans mass: $m_\mathrm{gas} > m_\mathrm{jeans}$
	\item Star particle mass is greater than a threshold mass
\end{enumerate}

Once these conditions are fulfilled, a star particle will be inserted with a mass of 
\begin{equation}\label{eq:mstar}
m_* = m_{\mathrm{gas}} \times f_*,
\end{equation}where $m_{\mathrm{gas}}$ is the gas mass in the cell and $f_*$ is the parameter governing the conversion efficiency of the gas mass within a cell into a star particle. It is placed in the center of the cell with the same peculiar velocity as the gas in the cell. An equivalent mass of gas to that of the star particle is then removed from the cell to ensure mass conservation.

Despite the instantaneous creation of the star particle, the feedback happens over a longer period of time, to mimic the gradual process of star formation. Within each timestep ($dt$), the amount of star-forming gas is given by
\begin{equation}
\begin{split}
\label{eq:mform}
m_{\mathrm{form}} &= m_*\left[\left(1+\frac{t-t_0}{t_\mathrm{dyn}}\right)\mathrm{exp}\left(-\frac{t-t_0}{t_\mathrm{dyn}}\right)\right.
\\&-\left.\left(1+\frac{t+dt-t_0}{t_\mathrm{dyn}}\right)\mathrm{exp}\left(-\frac{t+dt-t_0}{t_\mathrm{dyn}}\right)\right],
\end{split}
\end{equation} 
where $t_0$ and $t$ are the creation time of the star particle and current time in the simulation respectively. Through this implementation, according to Equation \ref{eq:mform}, the rate of star formation increases linearly and peaks after one dynamical time before declining exponentially \citep{2011ApJ...731....6S}. The feedback injects mass, energy, and metals per timestep and continues until 12 dynamical times after its creation. 

As we are only interested in calibrating the amount of feedback energy injected, we will focus on the following to describe how it is being calculated. Interested readers can refer to Section 2.2 in \citet{2020MNRAS.497.5203O} for the discussion on mass and metals. The amount of feedback energy injected is regulated by $\epsilon$
\begin{equation}
E_\mathrm{feedback} = m_\mathrm{form} \times c^2 \times \epsilon
\end{equation} where $\epsilon$ and $c$ are the feedback efficiency and speed of light respectively. As a reference, for $\epsilon = 10^{-5}$ \citep{1992ApJ...399L.113C}, $\mathrm{10^{51}\,erg}$ of energy is injected per approximately $56\,\mathrm{M_\odot}$ of stars formed.

Lastly, $r\_s$ allows the user to define the volume into which the feedback energy is injected into. For example, $r\_s = 1\_1$ refers to the six adjacent cells to the star particle. Increasing this parameter to $1\_2$ or $1\_3$ increases the number of cells and hence volume. However, it decreases the amount of feedback energy per cell (refer to Section 2.2.2 in \cite{2020MNRAS.497.5203O} for more details). To put it simply, $r\_s = 1\_1 \approx 63.5\,{\rm ckpc}^3$ and $r\_s = 1\_3 \approx 285.7\,{\rm ckpc}^3$. These are the values that will be of importance in later sections.

We are calibrating $f_*$, $\epsilon$ and $r\_s$ to the baryon makeup, specifically the ratio of stars and gas to the total baryonic mass of a MW-sized halo at $z=0$. The ratios are quantified in the same way to that presented in \citet{2010ApJ...708L..14M}. Using different methods to obtain the equivalent circular velocity, the authors determined the total mass budgets in a variety of galaxies. They used $r_{\rm 500}$, the radius which encloses a density 500 times the critical density of the universe, according to the cluster data used in the analysis.
Identical to \citet{2010ApJ...708L..14M}, we quantify the MW-sized halo with the ratio of expected baryons that are detected, 
\begin{equation}
f_{\rm d} = \frac{m_{\rm b}}{f_{\rm b} \times m_{500}},
\label{eq:fd}
\end{equation} and the ratio of stars, 
\begin{equation}
f_{\rm s} = \frac{m_*}{f_{\rm b} \times m_{500}},
\label{eq:fs}
\end{equation} where $m_{\rm b}$ and $m_{500}$ refer to the baryonic and total mass within $r_{\rm 500}$ respectively, and $f_b$ is the universal baryon fraction determined to be $0.17 \pm 0.01$ \citep{2009ApJS..180..330K}. 

We then compare these values with that obtained from the analytical fit to \citet{2010ApJ...708L..14M}, given by equations 10 to 13 and Figure 2 of \citet{2020MNRAS.497.5203O}. The difference between the simulated and observed properties is quantified by $\left|\Delta f_{\rm d}\right|$ and $\left|\Delta f_{\rm s}\right|$. To obtain a perfect match between the simulated and observed MW-sized galaxy, $\Delta f_{\rm d}$ and $\Delta f_{\rm s}$ should be zero.

\subsubsection{Metallicity distribution simulation} \label{sec:ej_setup}

The aim of this part of the study is to use machine learning methods to calibrate the parameters required to obtain an agreement between a particle-based code ({\tt GIZMO} \citep{2015MNRAS.450...53H}) and a grid-based code ({\tt Enzo}) at a level consistent with that in \citet{2021ApJ...917...12S}.
We initialize an ideal simulation of an isolated Milky Way-sized disk galaxy with the initial conditions from the {\it AGORA} Project \citep{AGORA2016}. 
It models a dark matter halo following the Navarro-Frenk-White (NFW) profile \citep{1997ApJ...490..493N}, an exponential disk of stars and gas and a stellar bulge with a Hernquist profile \citep{1990ApJ...356..359H}. \footnote{The {\it AGORA} initial conditions are publicly available at \url{http://sites.google.com/site/santacruzcomparisonproject/blogs/quicklinks/}.} 
For detailed properties, readers are directed to Table 1 in \citet{2021ApJ...917...12S}. 
They will be evolved with {\tt Enzo} and {\tt GIZMO} using the pressure-energy smoothed particle hydrodynamics \citep[hereafter PSPH;][]{Hopkins2013}. 

Instead of the default setup in {\it AGORA}, following the findings of \citet{2021ApJ...917...12S}, we insert additional gas particles in the halo region with $n_{\rm H}=10^{-6}\,{\rm cm}^{-3}$ for the initial conditions of the simulation with {\tt GIZMO}. 
The mass of these particles matches that of those in the disk region, $m_{\rm gas, IC}=8.593\times10^{4}\,M_\odot$, translating to a total of 4,000 gas particles in the halo. 
They are given an initial metallicity and temperature of $Z_{\rm halo}=10^{-6}\,Z_{\rm disk}$ and 10$^{6}$ K respectively. 
This modification is necessary to reconcile the metal transport behavior between a grid-based and a particle-based simulation code. 

We apply the cubic spline kernel \citep{1989ApJS...70..419H} for the softening of the gravitational force with the desired number of neighboring particles $N_{\rm ngb}=32$. 
We also adopt the Plummer equivalent gravitational softening length $\epsilon_{\rm grav}$ of 80 pc and allow the hydrodynamic smoothing length to reach a minimum of $0.2\epsilon_{\rm grav}$. 
Radiative cooling, star formation, and feedback will follow the prescription in \citet{AGORA2016}.

The {\tt Enzo} simulation is the baseline for comparison to the {\tt GIZMO} simulation. 
We vary the diffusion coefficient, $C_{\rm d}$, and the thermal feedback energy budget in the {\tt GIZMO} simulation. 
$C_{\rm d}$ is used in the metal diffusion scheme in \citet{2018MNRAS.480..800H} and  \citet{Escala2018}, based on the Smagorinsky-Lilly model \citep{Smagorinsky1963, Shen2010}. 
It determines the effect of the subgrid diffusion influenced by the velocity shear between the particles, with the assumption that the local diffusivity is related to the velocity shear and resolution scale (refer to Section 2.3.3 in \citet{2021ApJ...917...12S} for more details). 
On the other hand, the amount of thermal feedback energy impacts the extent to which the metals are distributed and transported away from the central region of the galaxy. 
The halo metal mass ($M_{\rm metal}$) and gas mass of metallicity between $0.00885\,Z_\odot$ and $0.0115\,Z_\odot$ ($M_{\rm gas, Z\sim0.01}$) after $500\,{\rm Myr}$ of evolution will be compared between the two simulations. We choose to use $M_{\rm gas, Z\sim0.01}$ because the gas within this metallicity range is found to be highly sensitive to $C_{\rm d}$ as illustrated in Figure 9 of \citet{2021ApJ...917...12S}. This small range of metallicity is chosen arbitrarily to keep the output required for the training of the machine simple.    

\subsection{Machine learning architecture and approach} \label{sec:hj_setup}

\subsubsection{Neural density estimator}\label{sec:hj_neural}
Given finite computational resources and time, it is impossible to fully explore the parameter space of the subgrid physics in a numerical simulation as described in the previous sections. However, this step is necessary and vital for the calibration of the parameters in order for the simulation to produce results matching observations. We attempt to circumvent this problem by applying machine learning methods. While most machine learning methods require large data, we decide to utilize a neural density estimator, which makes use of a small number of data by representing the recommendation with a probability distribution. The values of the recommendation are then obtained by randomly sampling the output distribution.

We use a mixture density network \citep{bishop1994mixture} as the neural density estimator. This model represents the distribution with a set of Gaussian using the estimated mean, standard deviation, and probability of each Gaussian as an output to the given input. To train the network, we use the cross-entropy loss function

\begin{equation}
\mathcal{L} =-{\rm log}\sum_{i=1}^{m}p_i(x)N(y|\mu_{i}(x),\sigma_{i}(x))
\end{equation}
where $x$ and $y$ is the input and output, respectively. $m$ is the number of Gaussian, $N$ is the probability density function of the Gaussian distribution, $\mu$ and $\sigma$ is the mean and standard deviation of the Gaussian distribution respectively. The aim is to minimize $\mathcal{L}$ while training the neural network by fitting the mixture of Gaussian distribution to maximize the likelihood on the given data.

To obtain the recommendation of the properties, the subgrid parameters in different studies serve as input to the neural network, giving a mixture of Gaussian distributions and their respective probability. One of the proposed distributions is selected via Gumbel sampling \citep{2016arXiv161101144J, 2016arXiv161100712M} with the output probability. The selected Gaussian distribution is then randomly sampled to obtain the recommendation. As such, the number of Gaussian distributions is one of the hyper parameters of the neural network, which will be discussed in Section \ref{sec:hj_results}. We also look at the impact of the number of initial data and the number of additional data per iteration on the performance of the machine.

\subsubsection{Active learning}
\begin{figure*}
\centering
\includegraphics[scale=0.6]{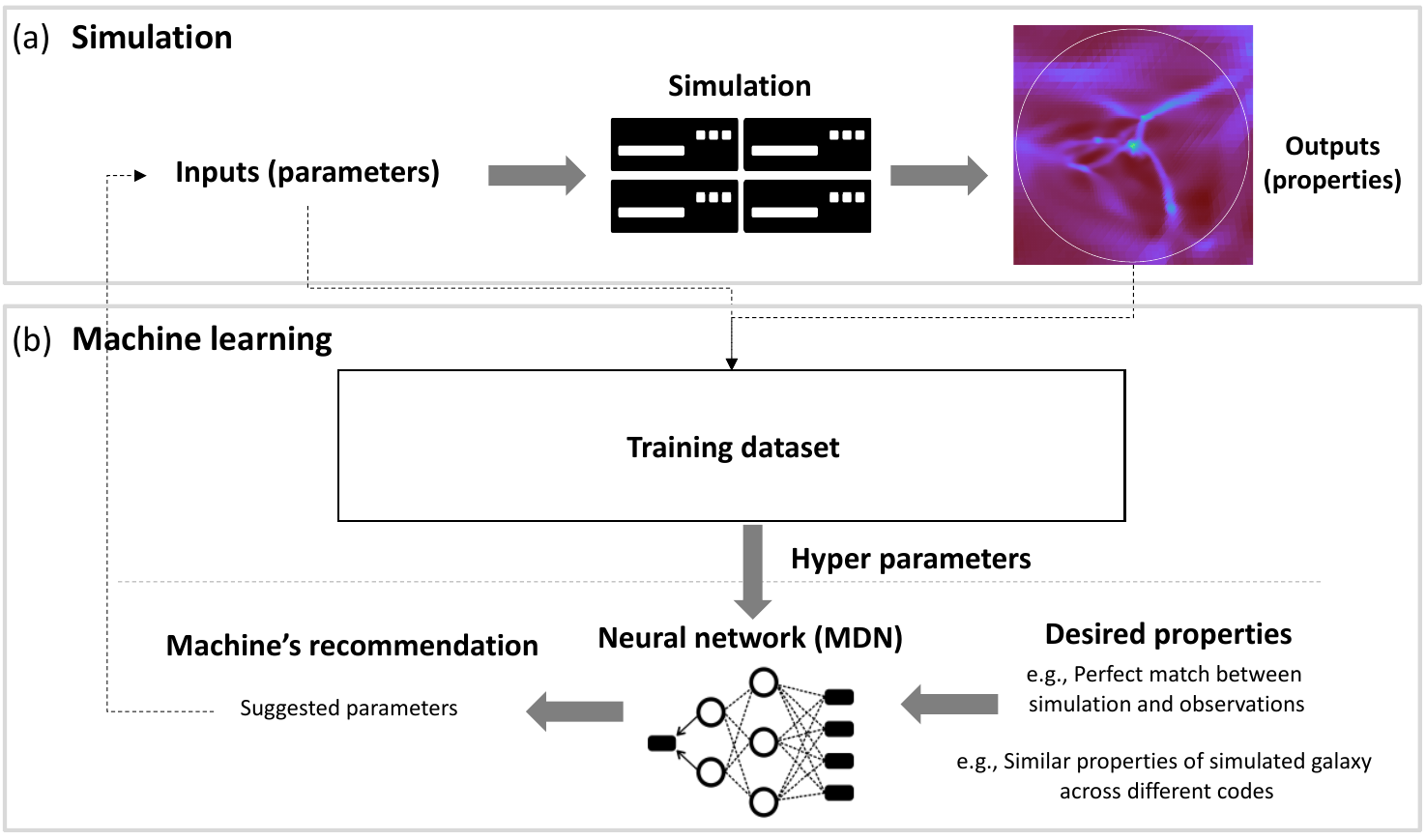}
\caption{Overview of the calibration using active learning. (a) Simulation part for the forward path. The simulation is performed to obtain parameters from given properties. (b) Machine part for backward path. The results of the simulation are added to the training dataset for the neural network. The trained network then recommends a new set of parameters to match observation. This recommendation is passed to the next forward path. The process terminates when simulated parameters match the desired parameters within a certain limit.}
\label{fig:hj_setup}
\end{figure*}

In this section, we will describe the architecture of the machine used as shown in Figure \ref{fig:hj_setup}. We apply active learning, which is a method of machine learning designed to reduce data generation through the interaction between data generation and machine. It is an iterative process that focuses attention on regions of input space that produce the desired output. In this study, we utilize active learning to minimize the total number of simulations required for the calibration, either to match observations (see Section \ref{sec:bk_setup}) or to reconcile the results from different simulation codes (see Section \ref{sec:ej_setup}). 

The pipeline consists of a forward and a backward path. Starting with an initial set of data, which includes both the input subgrid parameters and simulated output properties, the neural network is trained to provide a set of recommended subgrid parameters according to Section \ref{sec:hj_neural}. The desired output is given to the machine for the generation of the recommendation. For example, in Section \ref{sec:bk_setup}, the ideal case is a perfect match between the simulated and observed properties, i.e.,  $\left|\Delta f_{\rm d}\right| = 0$ and $\left|\Delta f_{\rm s}\right| = 0$. This process constitutes the backward path, in which the machine provides recommended values of subgrid parameters for the simulation. 

We then proceed with the forward path. A simulation is carried out with this set of proposed parameters until its desired endpoint. The simulated properties are then noted and checked against the criteria described in the previous section in order to define them as a good match. In the case that the simulation is not, both the input parameters and the output properties are consolidated into a training dataset. Having an additional data point than before, the training data is fed into the backward path again to produce another set of recommendations. This process is repeated until we fulfill the criteria of a good match. For the calibration of feedback parameters in cosmological simulation, a good match is obtained when $\left|\Delta f_{\rm d}\right| < 0.0577$ and $\left|\Delta f_{\rm s}\right| < 0.0748$ at $z=0$, given by \citet{2020MNRAS.497.5203O}. On the other hand, for the metal distribution simulations, if $M_{\rm metal}$ and $M_{\rm gas, Z\sim0.01}$ from the {\tt GIZMO} simulation falls within 10\% of $1.09\times10^6\,M_\odot$ and $2.04\times10^6\,M_\odot$ respectively after $500\,{\rm Myr}$, it will be classified as a good match. These values of comparison are obtained from an {\tt Enzo} simulation.

The weights of the neural network can be adjusted in two different ways. The first is a reset after every iteration through the forward and backward path while the other is an adjustment of the weights via each iteration. Due to the unique nature of the problem, which is the limited number of data and the high costs of adding every additional data point, we have to prevent a situation of getting stuck in a local minimum as a result of the low number of data. That is why we choose to flush the weights after every iteration.

In the following sections, we will discuss the determination of the hyper parameters of the machine using a projectile motion problem. It is structured to mimic the number of inputs and outputs of the problem described in Section \ref{sec:bk_setup}, which is three and two respectively. We can get an estimate of the suitable values by testing this problem with a well-defined solution. 

\section{Results and discussion} \label{sec:results}
\subsection{Calibration of hyper parameters using projectile motion with air resistance} \label{sec:hj_results}

In this section, we explore a projectile motion under the influence of air resistance in order to select the hyper parameters of the machine. Ideally, the parameters should be calibrated with a large amount of data that is identical to the problem. However, in a realistic feedback calibration, it is prohibitively expensive to run numerous simulations. Hence, we apply the machine to a well-posed problem: projectile motion with air resistance, in order to explore the hyper parameter space. 
We set up the machine to possess the same number of inputs and outputs for the neural network to the calibration of feedback parameters in Section \ref{sec:bk_setup}. The maximum vertical height and horizontal distance of the projectile motion (two outputs) are calculated from the initial velocity, drag coefficient, and launch angle of the object (three inputs). Using the target of a maximum distance and height of $2.3\,$m and $0.6\,$m respectively, we ask the machine to provide recommendations of the inputs required for this projectile motion. With this setup, we tune the number of initial training data, the number of additional data per iteration, and the number of Gaussian distributions on the proximity of the outputs from the recommended inputs to the target.

The calculation is performed using the Euler forward method for the differential equation with a maximum of 100 iterations. The air resistance or Stokes drag applies for very low speed in air and it is linearly increasing with velocity,
\begin{equation} 
\ddot{\vec{x}} = -b \dot{\vec{x}} +\vec{g}
\end{equation} where $\vec{x}$ denotes a position vector, $\cdot$ denotes derivative with respect to time, $b$ denotes the drag coefficient, and $\vec{g}$ is gravitational acceleration vector. Stokes drag is included to increase the complexity of the problem. The maximum height and distance reached by the projectile motion according to the inputs are then recorded. We allow $b$, the initial velocity and angle of the projectile to be randomly picked between 0 and 1, $0\,{\rm m/s}$ to $10\,{\rm m/s}$ and 0$^{\circ}$ to 90$^{\circ}$, respectively. The recommended velocity and $b$ must be more than zero for the problem to be realistic. It is noted that multiple combinations will produce the desired projectile motion but an example set of inputs to achieve a distance of $2.3\,{\rm m}$ below a height of $0.6\,{\rm m}$ is a launch velocity of $5\,{\rm m/s}$ at an angle of $45\,^{\circ}$ with $b=0.25$.

For the initial training set, we vary the number of data, using intervals of [5, 10, 20, 30, 40, 50]. The number of Gaussian distributions used by the machine and the number of additional data introduced per iteration are fixed to 10 and 1 correspondingly. We then experimented with the number of Gaussian distributions from 5 to 50 in intervals of 5 while fixing the number of initial data to 5 and the number of additional data per iteration to 1. Lastly, we increase the number of additional data per iteration from 1 to 2 while keeping the number of Gaussian distribution and initial data to 20 and 5 respectively. This range of parameters is considered small as compared to typical machine learning problems because it is intended to replicate the situation faced when calibrating a simulation. 

The training is performed for 2,000 epochs using the ADAM optimizer \citep{kingma2014adam} with a learning rate scheduler (ReduceLROnPlateau; initial learning rate: 1e$^{-3}$, decay factor: 0.5, patience: 500 epochs, and relative threshold: 1e$^{-3}$). To reduce the effects of randomness, the experiment of each comparison is repeated 50 times and the weights initialization is performed with a fixed random seed. The initial data configuration is also fixed. Python and Pytorch library \citep{paszke2019pytorch} is used for programming and an Nvidia TITAN Xp GPU (Nvidia Corp., Santa Clara, CA) is utilized.

\begin{figure}
\centering
\includegraphics[width=\linewidth]{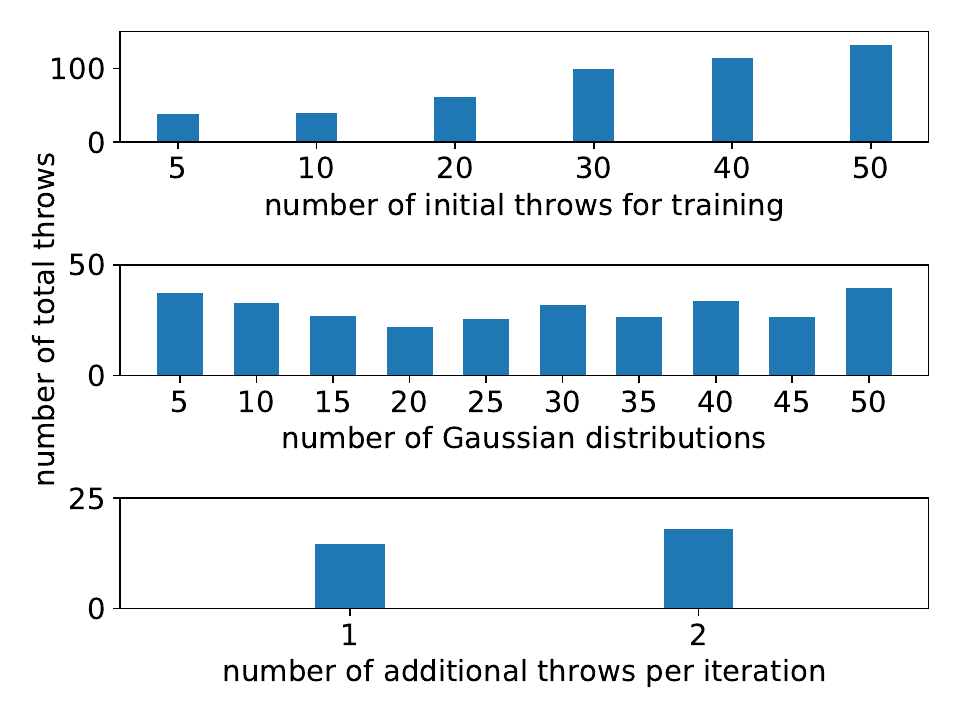}
\caption{Optimisation of the hyper parameters used in the machine. The top, middle and bottom panel corresponds to the total number of throws required to obtain $\Delta_{\rm total}\leq5\%$ in relation to the number of initial throws used for training, the number of Gaussian distributions used by the machine and number of additional throws per iteration respectively. Only one parameter is allowed to vary per experiment. We find that five initial throws, 20 Gaussian distributions and one additional throws per iteration yields the minimum number of total throws.}
\label{fig:hyperpara}
\end{figure}

We analyze the exploration of the hyper parameters required to reach within 5\% of the total error $\left(\Delta_{\rm total}\right)$. It is calculated using the target distance $\left({\rm dist}_{\rm target}=2.3\,{\rm m}\right)$ and target height $\left(h_{\rm target}=0.6\,{\rm m}\right)$: $\sqrt{\left(\Delta h/h_{\rm target}\right)^2 + \left((\Delta {\rm dist}/{\rm dist}_{\rm target}\right)^2}$ where $\Delta h$ and $\Delta {\rm dist}$ is the difference between the height and distance of each throw and their respective target. This optimisation of hyper parameters is shown in Figure \ref{fig:hyperpara}. We find that using five initial throws yield the smallest amount of additional throws required to reach within 5${\%}$ error (21.9 ± 20.3). The total number of throws increases proportionally to the number of initial throws. On the other hand, the number of throws required by varying the number of Gaussian distributions does not show a clear tendency as indicated in the second row of Figure \ref{fig:hyperpara}. Among them, using 20 Gaussian distributions requires the least amount of additional throws to reach within 5${\%}$ error (14.7 ± 5.1). Lastly, we vary the number of additional throws per iteration. While it is easy to increase the training data by a large amount, we have to remember that each throw is equivalent to a simulation is the problem we are interested in. Hence, the increment is limited to one or two per iteration, shown in third row of Figure \ref{fig:hyperpara}. We find that increasing the training set by one data per iteration yielded a slightly better performance (14.7 ± 5.1) than two data per iteration (18.2 ± 6.4). 

\begin{figure}
\centering
\includegraphics[width=\linewidth]{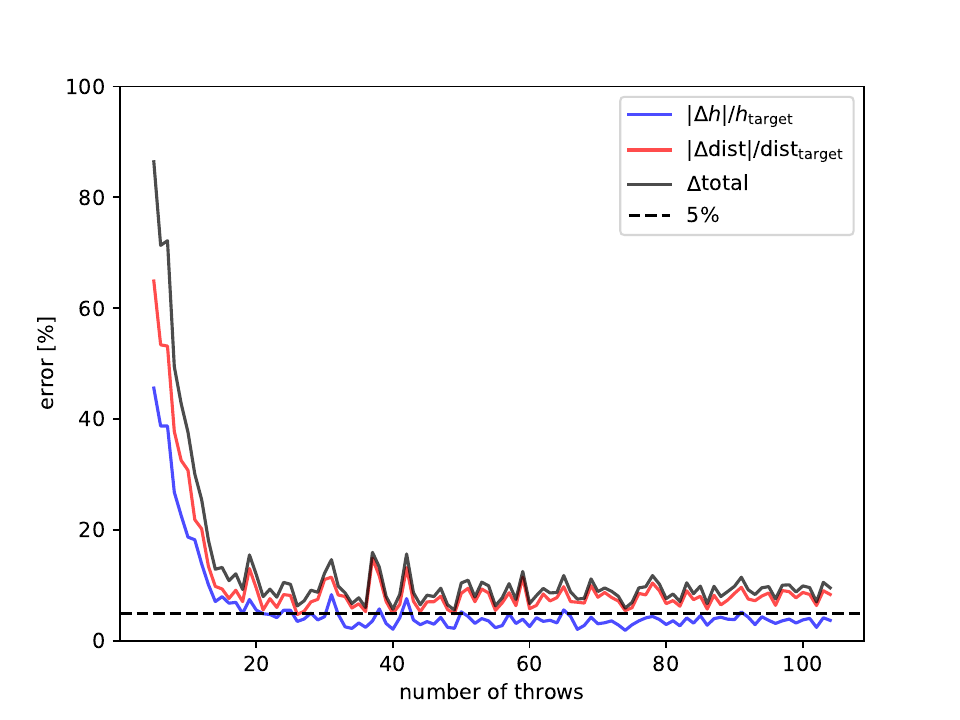}
\caption{Evolution of the mean $\Delta h$ (blue), $\Delta {\rm dist}$ (red) and $\Delta {\rm total}$ (black) with the total number of throws in 100 experiments. The dotted line represents the target error range of 5\%. In general, the error of all three quantities decreases as the number of throws increases, which indicates the success of the machine to provide meaningful recommendations.}
\label{fig:mse}
\end{figure}

With this setup of 5 initial throws for training, 20 Gaussian distributions for the machine, and one additional throw per iteration, we run the machine for 100 times with different initial training sets to gauge its performance. Figure \ref{fig:mse} shows the evolution of the mean error in the distance, height and combination of both with the number of throws. It is evident that as the number of throws increases, the deviations from the target decreases. In other words, the recommendation from the machine is improving the accuracy with each additional throw. Also, the throw reaches within 5\% of the target height quicker than the distance, perhaps due to the relatively simpler equation of motion in the y-axis of the throw.

After calibrating the hyper parameters, the machine is then applied to the astrophysical simulations with different strategies of obtaining the recommendation. We will illustrate the discrepancy in the machine's performance with a full exploration (see Section \ref{sec:bk_results1}), explore versus exploit (see Section \ref{sec:bk_results2}), human intervention strategy (see Section \ref{sec:ej_results_human}) and modifying the hyper parameters (see Section \ref{sec:ej_results_machine}. These strategies will result in different behaviour and performance of the machine. We can see some indications from the behaviour of the $\Delta {\rm total}$ as it does not continue to reduce. It means that there is no further exploration of the parameter space around the optimal. It will be discussed and addressed in later sections.

\subsection{Calibration of a star formation and feedback model: explore} \label{sec:bk_results1}
Together with the machine setup described at the end of the previous section, we apply it to calibrate the star formation and feedback model described in Section \ref{sec:bk_setup}. We tune the $f_*$, $\epsilon$, and $r\_s$ to produce the observed $f_{\rm d}$ and $f_{\rm s}$ of a MW-sized halo at $z=0$. The initial training data, which is designed to maximize the coverage in parameter space is shown in Table \ref{tab:bk_ini1}.

\begin{table}
	\caption{Initial set of training data for calibrating the baryon makeup in a MW-sized halo with a {\tt Enzo} simulation. This table includes the star formation efficiency ($f_*$), volume in which the feedback is injected into ($r\_s$), feedback efficiency ($\epsilon$), which are the inputs to the machine. The difference between the resulting and observed $f_{\rm d}$ and $f_{\rm s}$ (see equations \ref{eq:fd} and \ref{eq:fs}) are the outputs of the machine. It is designed to provide recommendations based on the minimisation of $\left|\Delta f_{\rm d}\right|$ and $\left|\Delta f_{\rm s}\right|$. Refer to Section \ref{sec:bk_results1} for more information.}
	\label{tab:bk_ini1}
	\centering
	\begin{tabular}{|p{.15\columnwidth}|p{.15\columnwidth}|p{.15\columnwidth}|p{.15\columnwidth}|p{.15\columnwidth}|}
		\hline
		\multicolumn{5}{|c|}{{\tt Enzo} training data}\\
		\hline
		\multicolumn{3}{|c|}{Inputs (paramters)} & \multicolumn{2}{c|}{Outputs (properties)}\\
		\hline
		$f_*$ & $r\_s$ & $\epsilon$ & $\left|\Delta f_{\rm d}\right|$ & $\left|\Delta f_{\rm s}\right|$\\ 
		\hline
		5.0e-05& 2$\_$6 & 0.1 & 0.31 & 0.11 \\ 
        \hline
        5.0e-06 & 2$\_$6 & 0.2 & 1.11 & 1.05 \\
        \hline  
		2.5e-04 & 1$\_$3 & 0.6 & 0.06 & 0.18 \\
		\hline
		1.0e-06 & 1$\_$3 & 0.9 & 1.18 & 1.13 \\
		\hline
		2.0e-05 & 1$\_$1 & 0.9 & 0.26 & 0.04 \\
		\hline
		\multicolumn{5}{|c|}{Values needed for a perfect match between simulation and observations}\\
		\hline
		- & - & - & 0 & 0\\
		\hline
	\end{tabular}
\end{table}

\begin{figure*}
	\begin{subfigure}{0.49\textwidth}
		\includegraphics[width=\linewidth]{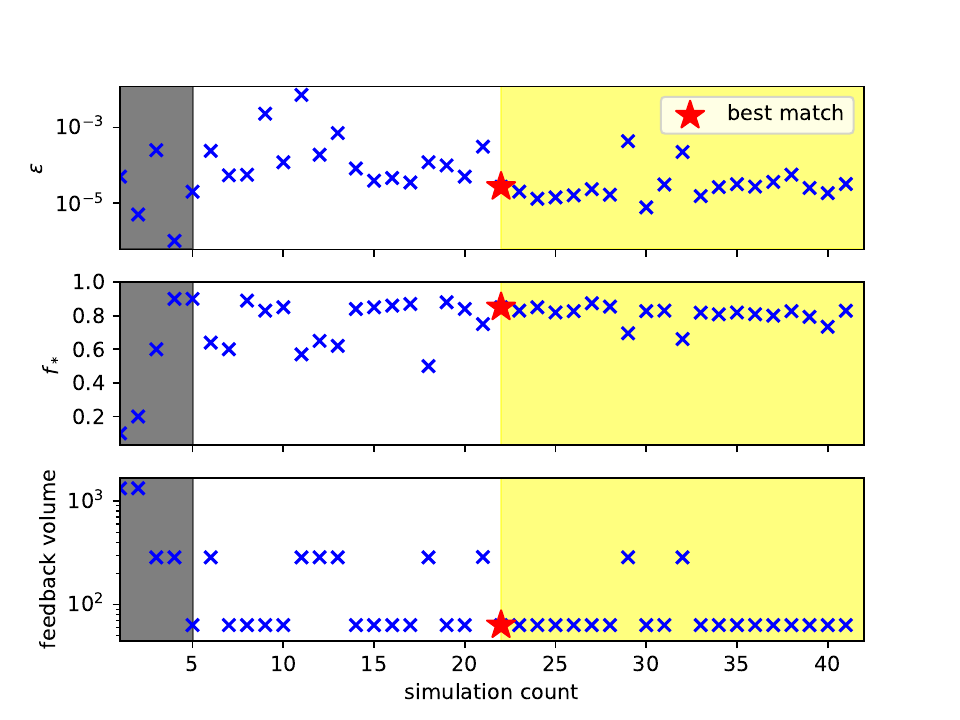}
		\label{fig:hj_parameter}
	\end{subfigure}
	\begin{subfigure}{0.49\textwidth}
		\includegraphics[width=\linewidth]{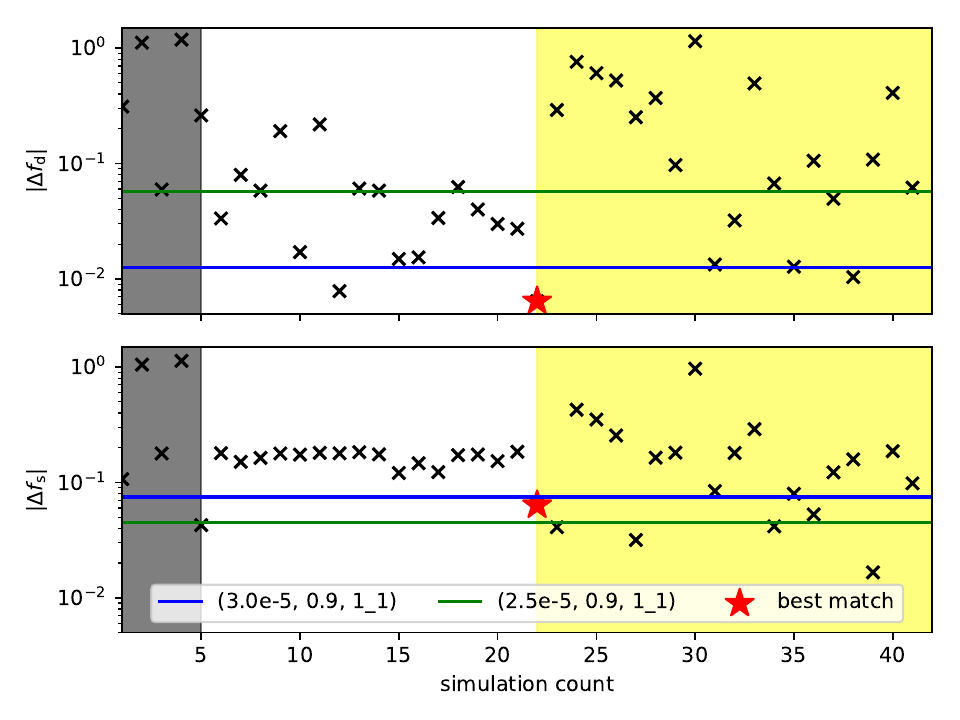}
		\label{fig:hj_match}
	\end{subfigure}
	\caption{Iteration of simulation parameters recommended by the machine (left) and the corresponding outputs from the simulations (right) using the explore and exploit strategy. The blue cross in the left figure represents the $\epsilon$, $f_*$ and $r\_s$ in their respective panel. The black cross, blue and green lines in the right panel represents the output of each simulation, output of (3.0e-5, 0.9, 1\_1) and (2.5e-5, 0.9, 1\_1) from \citet{2020MNRAS.497.5203O} respectively. The top and bottom rows in the right panel corresponds to the evolution of $\left|\Delta f_{\rm d}\right|$ and $\left|\Delta f_{\rm s}\right|$. In both figures, we have indicated the initial training set consists of the first five data points within the black shaded area and the yellow shaded region illustrates the behavior of the outputs from the recommendations of the machine after the best match (red star) is found. A good match (2.7e-5, 0.85, 1\_1) is identified after 17 additional simulations, which is more than a factor of three improvements as compared to \citet{2020MNRAS.497.5203O}. It also yields a $\left|\Delta f_{\rm d}\right|$ which is better than previous work while retaining an acceptable $\left|\Delta f_{\rm s}\right|$, an in-between of those in \citet{2020MNRAS.497.5203O}. We will discuss the progression of the recommendations in Section \ref{sec:bk_results1}.}
	\label{fig:hj_results}
\end{figure*}

Instead of the raw $f_{\rm d}$ and $f_{\rm s}$, we use $\left|\Delta f_{\rm d}\right|$ and $\left|\Delta f_{\rm s}\right|$ (see Section \ref{sec:bk_setup}) as the input of the neural network, which means the desired value is zero in both cases. We also represent the $r\_s$ with the total volume into which the feedback energy is injected. Together with $f_*$, we scale both parameters logarithmically to capture slight changes in the values for the output of the neural network. It is important to note that we did not specify that the volume is discrete but the machine picks it up and starts recommending numbers around the discrete values of volume as the number of data points increases. By posing restrictions in the data provided to the machine in each iteration, the machine is able to adapt the recommendations accordingly. With 17 additional simulations, we identify a combination of parameters of (2.7e-5, 0.85, 1\_1) which simulated a MW-sized halo with $\left|\Delta f_{\rm d}\right| = 0.00642$ and $\left|\Delta f_{\rm s}\right| = 0.0636$ at $z = 0$.

Due to the exploratory nature of the machine, we can clearly identify the huge variations in the parameters recommended in the left panel of Figure \ref{fig:hj_results}. While originally designed to prevent getting trapped in a local minimum, the machine tends to move around the parameter space according to the Gaussian distributions recommended. In other words, unless it is a single distribution, the recommendation will not be consistent. This variation can be seen before and after the yellow region. Right before the best combination, the value of $\epsilon$ is still varying drastically. Even the feedback volume is alternating between two values before the best match is found. In contrast, \citet{2020MNRAS.497.5203O} narrows the variation down to a single parameter in the last steps. After the best match is found, the machine continues to fluctuate around the parameter space, which is not desirable. The machine should focus on these values and attempt to find a better match.

We get a different perspective on the recommendation of the parameters provided by the machine, in the output space (see right panel of Figure \ref{fig:hj_results}). As it is a direct comparison to the work by \citet{2020MNRAS.497.5203O}, we use their best-performing simulations to evaluating how well the machine performed. In the mentioned work, the authors found the best match with (3.0e-5, 0.9, 1\_1) and (2.5e-5, 0.9, 1\_1) which gave a $\left|\Delta f_{\rm d}\right|$ and $\left|\Delta f_{\rm s}\right|$ of 0.0748 and 0.0126 (blue line), and 0.0577 and 0.0447 (green line) respectively. These values are used as constraints under which the properties from our simulations must fall under. The $\left|\Delta f_{\rm d}\right|$ and $\left|\Delta f_{\rm s}\right|$ of our simulated halo are 0.00642 and 0.0636. While the $\left|\Delta f_{\rm s}\right|$ is comparable, $\left|\Delta f_{\rm d}\right|$ is an order of magnitude better than previous results, which means the composition of gas and stars is more accurate. However, the machine fails to continue exploring this region of parameter space (see left panel of Figure \ref{fig:hj_results}) and starts suggesting simulations with properties that deviate more from observations. Given the identification of a set of parameters able to reproduce observations, it is desirable that the machine explore this region of parameter space further to possibly locate a better match. For example, if a set of parameters yields properties just above the criteria of acceptance, further exploration around these parameters will be more efficient than trying a completely different region in parameter space.

Within 22 simulations, more than three times fewer simulations, we obtain a set of parameters that produced a MW-sized halo with a baryon makeup that is closer to observations than \citet{2020MNRAS.497.5203O}. This saving translates to an equal amount of computational resources. While acceptable due to our intent to fully explore the parameter space, the recommendations leading up to the best one are random. This behavior is even more undesirable after the best match is found because we hope for the machine to look at this region in greater detail to potentially find a better match.

\begin{figure*}
	\begin{subfigure}{0.49\textwidth}
		\includegraphics[width=\linewidth]{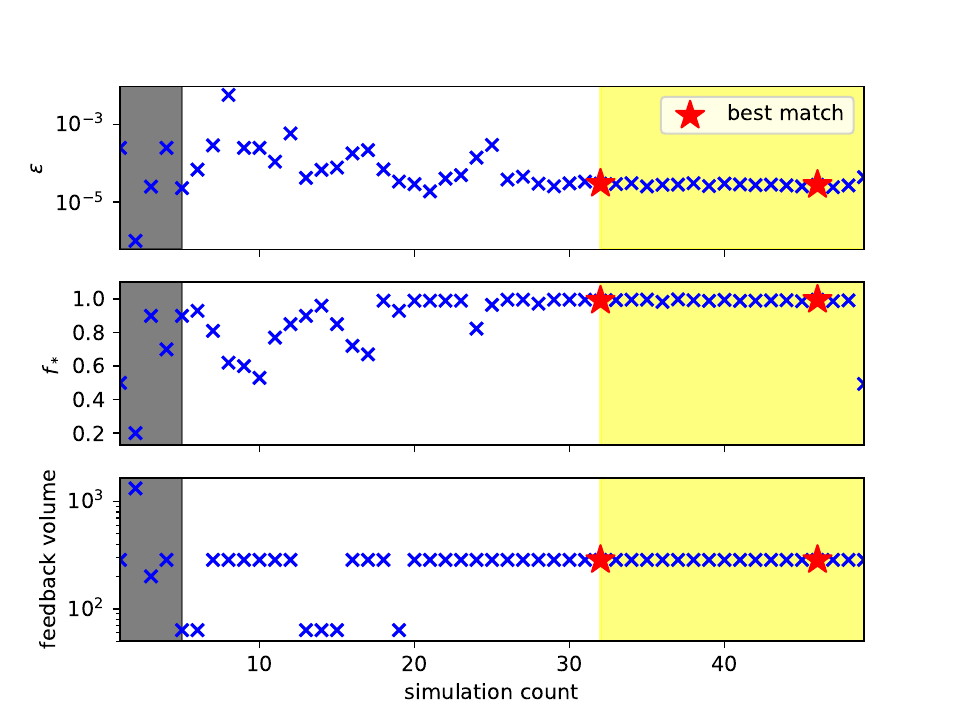}
		\label{fig:bk_parameter}
	\end{subfigure}
	\begin{subfigure}{0.49\textwidth}
		\includegraphics[width=\linewidth]{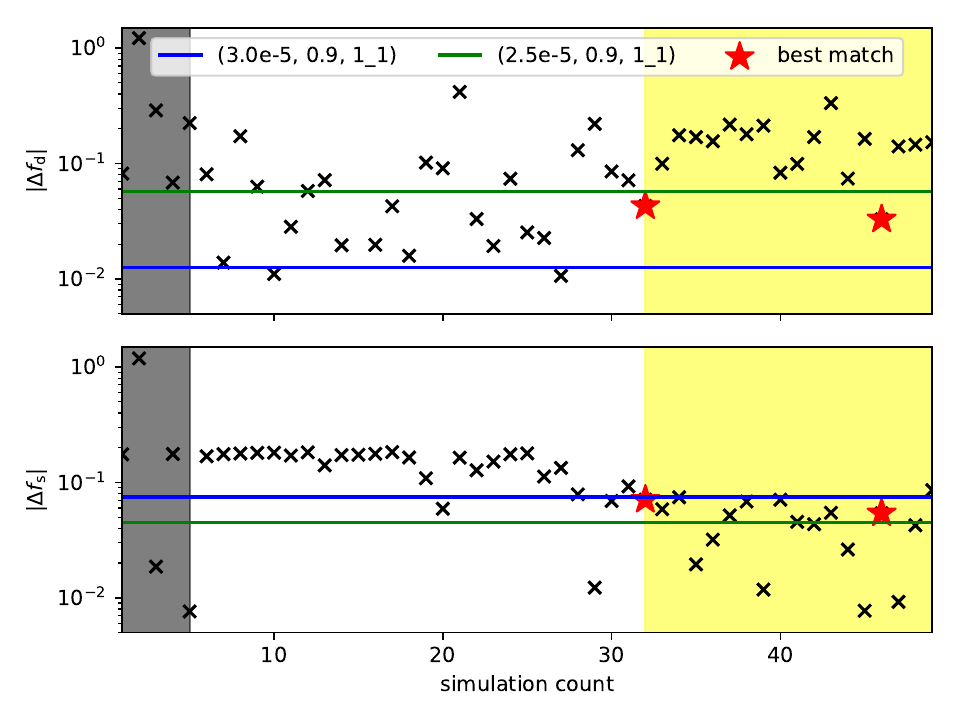}
		\label{fig:bk_match}
	\end{subfigure}
	\caption{Iteration of simulation parameters recommended by the machine (left) and the corresponding outputs from the simulations (right) using the explore and exploit strategy. The blue cross in the left figure represents the $\epsilon$, $f_*$ and $r\_s$ in their respective panel. The black cross, blue and green lines in the right panel represents the output of each simulation, output of (3.0e-5, 0.9, 1\_1) and (2.5e-5, 0.9, 1\_1) from \citet{2020MNRAS.497.5203O} respectively. The top and bottom rows in the right panel corresponds to the evolution of $\left|\Delta f_{\rm d}\right|$ and $\left|\Delta f_{\rm s}\right|$. In both figures, we have indicated the initial training set consists of the first five data points within the black shaded area and the yellow shaded region illustrates the behavior of the outputs from the recommendations of the machine after the best match (red star) is found. The best match is given by the machine by 32 simulations and a better one is found by 46 simulations. However, it takes more simulations to reach a worse agreement than Section \ref{sec:hj_results}. We discuss this discrepancy in Section \ref{sec:bk_results2}.}
	\label{fig:bk_results}
\end{figure*}

Though we obtained a better agreement between the simulated and observed properties of the MW-sized halo, it is still not possible to achieve a perfect match, i.e., $\left|\Delta f_{\rm d}\right| = 0$ and $\left|\Delta f_{\rm s}\right| = 0$. Given the inability to do so, we can consider increasing the number of parameters used to calibrate the Cen \& Ostriker model of star formation and feedback. Alternatively, there might be inherent limitations in the application of this model, at least within the resolution limits of our simulation. We aim to address the latter by changing the strategy used by the machine to explore the parameter space in detail.

\subsection{Calibration of a star formation and feedback model: explore \& exploit} \label{sec:bk_results2}

The calibration of feedback parameters with the novel method is successful in terms of the agreement and computational resources used but there is huge uncertainty in the exploration of the parameter space. This behavior is attributed to the exploratory nature of the strategy used by the machine. Out of all possible Gaussian distributions identified by the machine, the final set of parameters is drawn randomly within one of the distributions. Rather than ignoring the probability of each Gaussian distribution, we rank them accordingly. It focuses on drawing the recommendations from the distribution with the highest probability if it is higher than the next probable distribution by at least 20\%. On top of adjusting the strategy, we also change the set of initial training data to test the robustness of the machine. The five data points are picked to maximize coverage of parameter space, similar to that in Section \ref{sec:bk_results1}. We present them in Table \ref{tab:bk_ini2}.  

\begin{table}
	\caption{Initial set of training data for calibrating the baryon makeup in a MW-sized halo with a {\tt Enzo} simulation with a different strategy. This table includes the star formation efficiency ($f_*$), volume in which the feedback is injected into ($r\_s$), feedback efficiency ($\epsilon$) and the difference between the resulting and observed $f_{\rm s}$ and $f_{\rm d}$. The initial training data is different from that in Figure \ref{tab:bk_ini1} to showcase the robustness of the machine. Refer to Section \ref{sec:bk_results2} for more information.
	}
	\label{tab:bk_ini2}
	\centering
	\begin{tabular}{|p{.15\columnwidth}|p{.15\columnwidth}|p{.15\columnwidth}|p{.15\columnwidth}|p{.15\columnwidth}|}
		\hline
		\multicolumn{5}{|c|}{{\tt Enzo} training data}\\
		\hline
		\multicolumn{3}{|c|}{Inputs (paramters} & \multicolumn{2}{c|}{Outputs (properties)}\\
		\hline
		$f_*$ & $r\_s$ & $\epsilon$ & $\left|\Delta f_{\rm d}\right|$ & $\left|\Delta f_{\rm s}\right|$\\ 
		\hline
		2.5e-04 & 1$\_$3 & 0.5 & 0.08 & 0.18 \\ 
        \hline
        1.0e-06 & 2$\_$6 & 0.2 & 1.22 & 1.19 \\
        \hline  
		2.5e-05 & 1$\_$2 & 0.9 & 0.29 & 0.02 \\
		\hline
		2.5e-04 & 1$\_$3 & 0.7 & 0.07 & 0.18 \\
		\hline
		2.3e-05 & 1$\_$1 & 0.9 & 0.22 & 0.01 \\
		\hline
		\multicolumn{5}{|c|}{Values needed for a perfect match between simulation and observations}\\
		\hline
		- & - & - & 0 & 0\\
		\hline
	\end{tabular}
\end{table}

Adopting this new strategy allows the machine to sample the region that produced the best combination of parameters more thoroughly. While we assume an identical setup of hyperparameters, it is worthwhile to investigate it in a future work. As seen from the yellow region in the left panel of figures \ref{fig:hj_results} and \ref{fig:bk_results}, the variation in the parameter space decreased significantly in the latter. This difference suggests that by taking the probability into consideration, we can constrain the amount of randomness in the exploration of the parameter space. We find the values capable of reproducing observations to be $f_*=0.991$, $\epsilon=3.0\times10^{-5}$ and $r\_s=1\_3$, in contrast to Section \ref{sec:bk_results1} where $f_*=0.85$, $\epsilon=2.7\times10^{-5}$ and $r\_s=1\_1$. With a higher $f_*$, $\epsilon$ and $r\_s$, our current set of parameters is matching the effect of feedback energy injected per cell in Section \ref{sec:bk_results1}. It is consuming more gas per star particle creation and injecting correspondingly more energy in the increased volume. While the match is not perfect, it creates a MW-sized halo with a baryon makeup as good as \citet{2020MNRAS.497.5203O}.

The other difference is evident after the best set of parameters is determined, where the values stay close to the optimal combination of parameters as seen in the yellow region of the left panel in Figure \ref{fig:bk_results}. We have achieved our goal of better consistency from the recommendation of the machine. However, it takes more simulations to reach an equal or better agreement to \citet{2020MNRAS.497.5203O} than before. This difference in the number of required simulations can be attributed to this set of parameters being a local rather than a global optimum. We can validate this in the right panel of Figure \ref{fig:bk_results} where $\left|\Delta f_{\rm d}\right|$ and $\left|\Delta f_{\rm s}\right|$ by 32 simulations are more than the best match in the right panel of Figure \ref{fig:hj_results}.

\begin{figure*}
	\begin{subfigure}{0.49\textwidth}
		\includegraphics[width=\linewidth]{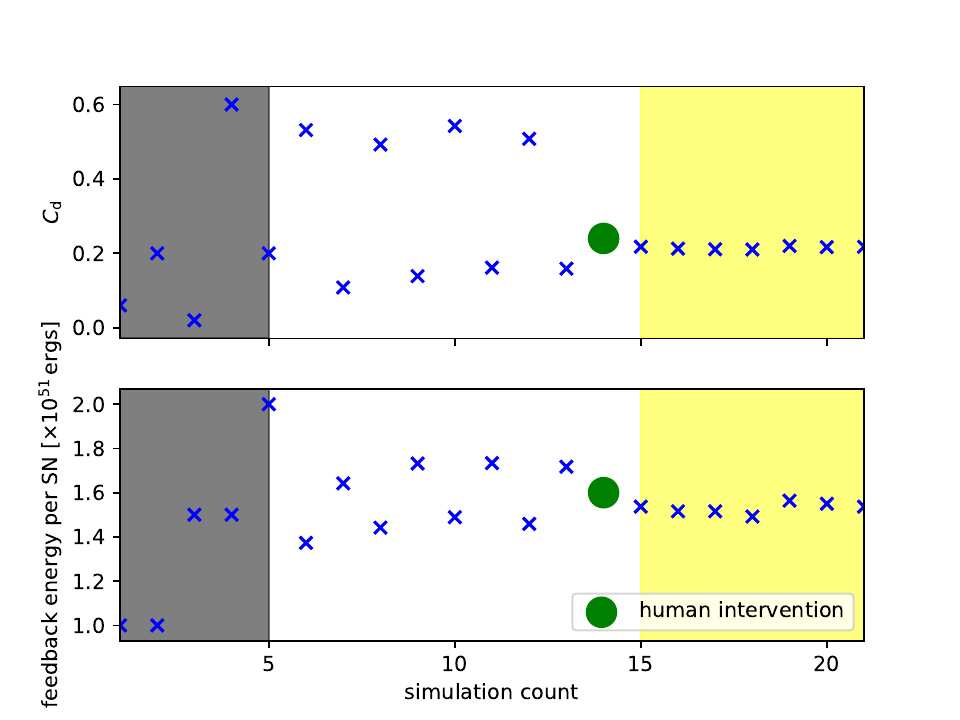}
		\label{fig:ej_parameter}
	\end{subfigure}
	\begin{subfigure}{0.49\textwidth}
		\includegraphics[width=\linewidth]{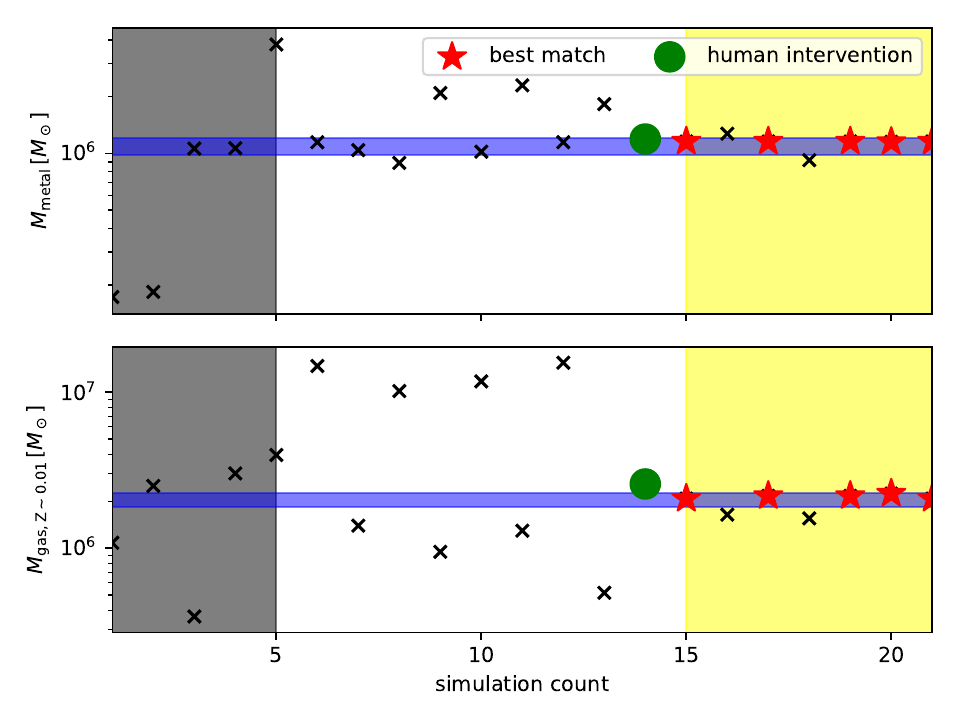}
		\label{fig:ej_match}
	\end{subfigure}
	\caption{Iteration of simulation parameters recommended by the machine (left) and the corresponding outputs from the simulations (right). The blue cross in the left figure represents the $C_{\rm d}$ and feedback energy per SN in multiples of $10^{51}\,{\rm ergs}$. The black cross, blue band, and red star in the right figure represents the output from the simulation, range of target values ($\pm 10\%$) from the {\tt Enzo} simulation, and the simulation that produced values within the blue bands in both panels respectively. In both figures, we have indicated the point of human intervention with the green dot, the initial training set consists of the first five data points within the black shaded area and the yellow shaded region illustrates the behavior of the outputs from the recommendations of the machine after the best match is found. We have decided to have a human-recommended input because of the oscillatory behavior of the parameters prior to the green dot. The details and impact of the human intervention will be discussed in Section \ref{sec:ej_results_human}}
	\label{fig:ej_results_human}
\end{figure*}

Between the right panel of figures \ref{fig:hj_results} and \ref{fig:bk_results}, we can see that there are two instead of one set of parameters that managed to simulate a MW-sized halo with a baryon makeup better than \citet{2020MNRAS.497.5203O}. We find a second match in 14 additional simulations (2.8e-5, 0.996, 1\_3) after the first (3.0e-5, 0.991, 1\_3), giving a decrease in $\left|\Delta f_{\rm d}\right|$ and $\left|\Delta f_{\rm s}\right|$, which is an improvement over the first. For the first match, $\left|\Delta f_{\rm d}\right| = 0.0428$ and $\left|\Delta f_{\rm s}\right| = 0.0708$ while the second match yields $\left|\Delta f_{\rm d}\right| = 0.0328$ and $\left|\Delta f_{\rm s}\right| = 0.0541$. Given the exploiting nature of the strategy in this section, this behavior is as expected at the costs of computational resources.

Combining the results of both sections \ref{sec:bk_results1} and \ref{sec:bk_results2}, we can conclude that the novel application of machine learning methods is successful in improving both the accuracy and computational resources spent on the calibration of the simulations. While Section \ref{sec:bk_results1} illustrated an effective strategy, it did not allow the possibility of further improving the agreement beyond the initial match. When we then take the possibilities of the Gaussian distributions into account, the machine provides a desirable behavior which is to focus on the parameter space that resulted in a good match. This agreement is however, poorer than before, likely due to the machine focusing on a local minima. If we include human analysis and steer the machine using human intervention, we can potentially further decrease the usage of computational resources. We will explore this strategy and the versatility of the machine in the following sections.

\subsection{Calibration of metal distribution in {\tt GIZMO} simulation with human intervention} \label{sec:ej_results_human}

To test the robustness of the architecture of the machine described in Section \ref{sec:hj_setup}, we extend its application to a different scientific problem explored in \citet{2021ApJ...917...12S}.
In this part of the study, we tried to identify values of the diffusion coefficient $C_{\rm d}$ and feedback energy budget in a {\tt GIZMO} simulation, which will reconcile with the transport behaviour in the {\tt Enzo} simulation. 
We compare $M_{\rm metal}$ and $M_{\rm gas, Z\sim0.01}$ defined in Section \ref{sec:ej_setup} after evolving for $500\,{\rm Myr}$ to that obtained in an {\tt Enzo} simulation.

\begin{table}
	\caption{Initial set of training data for calibrating the amount of metals in a MW-sized halo with a simulation using {\tt GIZMO} to that using {\tt Enzo}. This table includes the diffusion coefficient ($C_{\rm d}$, amount of feedback energy, and the resulting stellar mass, metal mass, and gas mass having a metallicity between $0.00885\,Z_\odot$ and $0.0115\,Z_\odot$. Refer to Section \ref{sec:ej_setup} for more information about the choice of parameters.}
	\label{tab:ej_ini}
	\centering
	\begin{tabular}{|p{.1\columnwidth}|p{.25\columnwidth}|p{.2\columnwidth}|p{.2\columnwidth}|}
		\hline
		\multicolumn{4}{|c|}{{\tt GIZMO} training data}\\
		\hline
		\multicolumn{2}{|c|}{Inputs (parameters} & \multicolumn{2}{c|}{Outputs (properties)}\\
		\hline
		$C_{\rm d}$ & Feedback energy [$\times10^{51}\,{\rm ergs\ per\ SN}$] & $M_{\rm metal}$ [$\times10^{5}\,M_\odot$] & $M_{\rm gas, Z\sim0.01}$ [$\times10^{5}\,M_\odot$]\\
		\hline
		0.006 & 1 & 1.73 & 10.81\\
		\hline
		0.02 & 1 & 1.84 & 25.07\\
		\hline
		0.002 & 1.5 & 10.61 & 3.64\\
		\hline
		0.06 & 1.5 & 10.66 & 30.13\\
		\hline
		0.02 & 2 & 37.86 & 39.56\\
		\hline
		\multicolumn{4}{|c|}{{Outputs (properties) from \tt Enzo} simulation}\\
		\hline
		- & - & 10.91 & 20.40\\
		\hline
	\end{tabular}
\end{table} 

With a similar setup using the hyper parameters described in Section \ref{sec:hj_results}, a training set comprising of five data sets is provided to the machine. While not comprehensive in the entire parameter space, they are chosen to maximize the coverage of the parameter space in \citet{2021ApJ...917...12S} as seen in Table \ref{tab:ej_ini}. The training data is illustrated in the black shaded region shown in the left panel of Figure \ref{fig:ej_results_human}.

An identical strategy to Section \ref{sec:bk_results1} is adopted. The machine is allowed to explore and recommend any values in the parameter space. However, from simulation number six to 14, the machine is oscillating between two prominent Gaussian distributions. It alternates between $C_{\rm d} \approx 0.05$ with ${\rm feedback\ energy\ per\ SN} \approx 1.4\times10^{51}\,{\rm ergs}$ and $C_{\rm d} \approx 0.014$ with ${\rm feedback\ energy\ per\ SN} \approx 1.7\times10^{51}\,{\rm ergs}$. Since this behaviour comes with a prohibitive cost computationally, we decide to try a new strategy, which is to introduce human intervention. The parameters we have introduced is motivated by the values of the $M_{\rm metal}$ and $M_{\rm gas, Z\sim0.01}$ in the right panel of Figure \ref{fig:ej_results_human}. It is determined to lie between the two oscillating peaks, $C_{\rm d} = 0.024$ with ${\rm feedback\ energy\ per\ SN} = 1.6\times10^{51}\,{\rm ergs}$.

\begin{figure*}
	\begin{subfigure}{0.49\textwidth}
		\includegraphics[width=\linewidth]{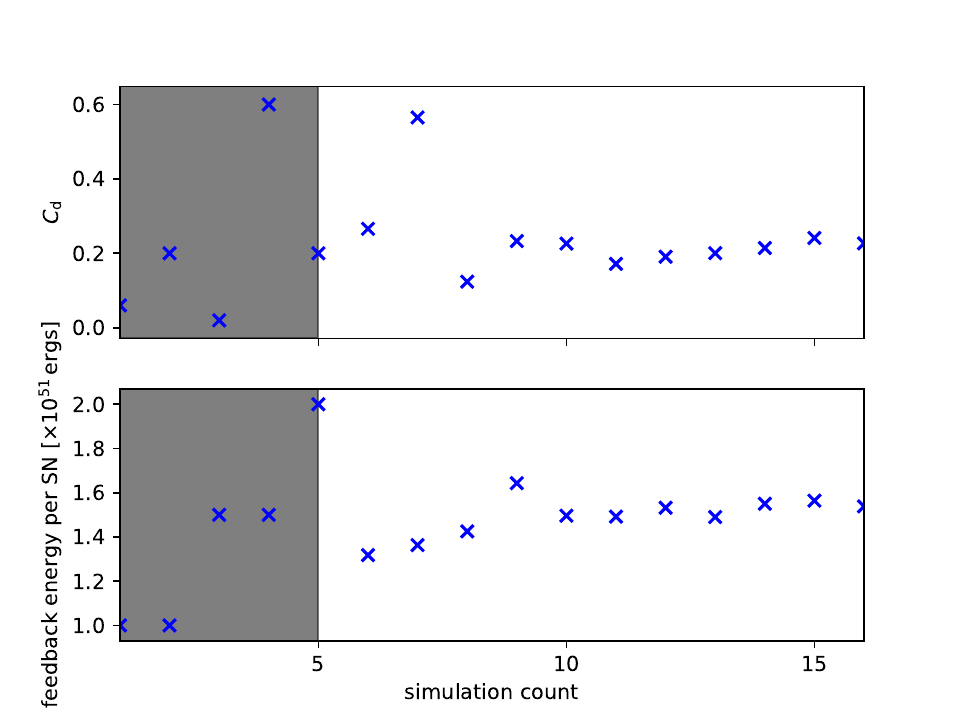}
		\label{fig:ej_parameter_mach}
	\end{subfigure}
	\hfill 
	\begin{subfigure}{0.49\textwidth}
		\includegraphics[width=\linewidth]{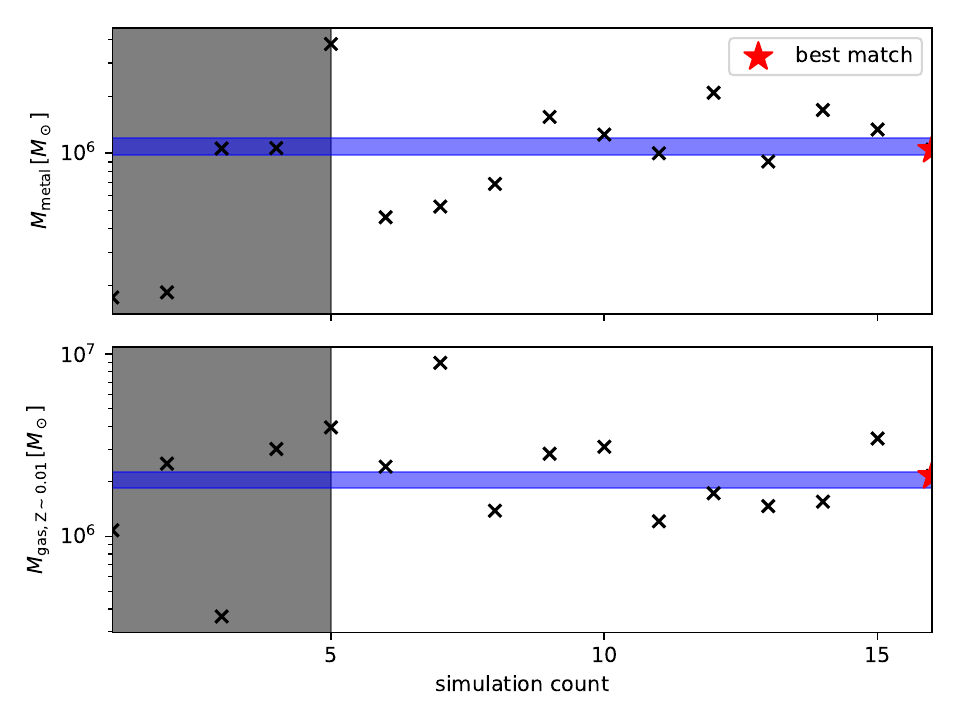}
		\label{fig:ej_match_mach}
	\end{subfigure}
	\caption{Similar to Figure \ref{fig:ej_results_human}. Instead of introducing a human-recommended input, we increased the number of Gaussians used in the machine learning algorithm to 100. The solution successfully prevented the occurrence of the oscillatory behaviour and pointed out the limitation of the hyper parameters used. These are discussed in Section \ref{sec:ej_results_machine}}
	\label{fig:ej_results_machine}
\end{figure*}

The initial purpose of the human intervention is to provide the machine with a {\tt GIZMO} simulation of a MW-sized galaxy that gave properties that resembles closer to that of the {\tt Enzo} simulation. We tried to steer the machine away from the oscillatory behavior by providing a better match. It worked, evident from the data after the green dot, even resulting in the best match being identified. After running the simulation with the human determined parameters, we obtained a MW-sized halo with $M_{\rm metal}$, $M_{\rm gas, Z\sim0.01}$ in the {\tt GIZMO} simulation to within 10\% of that in the {\tt Enzo} simulation. 

The recommendations that followed from the best match show consistency in both the parameter and properties space. They do not deviate far from each other or the target values from the {\tt Enzo} simulation. We use $C_{\rm d} \approx 0.0218$ and feedback energy per SN $\approx 1.54 \times 10^{51}\,{\rm ergs}$ in contrast to $C_{\rm d} = 0.02$ and feedback energy per SN $\approx 1.8 \times 10^{51}\,{\rm ergs}$ in \cite{2021ApJ...917...12S}. This discrepancy gives $M_{\rm metal}\approx1.16\times10^6\,M_\odot$ and $M_{\rm gas, Z\sim0.01}\approx2.08\times10^6\,M_\odot$ in our work as compared to $M_{\rm metal}\approx2.20\times10^6\,M_\odot$ and $M_{\rm gas, Z\sim0.01}\approx2.04\times10^6\,M_\odot$ in \citet{2021ApJ...917...12S}. While $M_{\rm gas, Z\sim0.01}$ is similar, $M_{\rm metal}$ is significantly different.

The results show that the machine recommended parameters produced a better agreement between the {\tt Enzo} and {\tt GIZMO} simulations, showcasing the improvement that is brought about through the usage of the machine learning methods. On top of this increment, the machine highlighted the sensitivity of $M_{\rm metal}$ to $C_{\rm d}$. With a higher feedback energy per SN, more metals are lost from the virial radius of the halo due to the rise in metal diffusion caused by the stronger outflow from the disk.

\subsection{Calibration of metal distribution in {\tt GIZMO} simulation by modifying hyper parameters} \label{sec:ej_results_machine}

From the experiment in Section \ref{sec:ej_results_human}, we corrected the oscillation of the recommendations obtained from the machine with a human intervention. While it has been proven to be effective in guiding the machine away from the undesirable behaviour, it requires a manual input which might differ from one to the other. In this section, we explore an approach that removes this uncertainty. 

Reviewing the architecture of the machine, we recognise that the set of hyper parameters is decided from a projectile motion problem. It is likely that this combination of parameters is not universal for all problems. Hence, we attempt to modify it to prevent the oscillatory behaviour. Specifically, we change the number of Gaussians only because the generation of more training data or more simulation per iteration will place a huge demand on computational resources. It is important to recognise that such an amount of resources might not be available readily.

Hence, we increased the number of Gaussians from 20 to 100 and illustrate the results in Figure \ref{fig:ej_results_machine}. The variation of the input parameters and output properties is shown on the left and right panel respectively. In comparison to Figure \ref{fig:ej_results_human}, we can see that the oscillations have been suppressed, indicating that the proposed change is beneficial. After a total of 16 simulations, with $C_{\rm d} \approx 0.227$ and feedback energy per SN $\approx 1.54\times10^{51}\,{\rm ergs}$, we obtain a simulated galaxy with $M_{\rm metal}\approx1.05\times10^6\,M_\odot$ and $M_{\rm gas, Z\sim0.01}\approx2.15\times10^6\,M_\odot$, within 10\% of the {\tt Enzo} simulation.

As discussed, with human intervention, there is a high degree of uncertainty to the choice of values used. Despite both experiments requiring a similar number of simulations to reconcile the properties of the simulated galaxy with a grid based and particle based code, increasing the number of Gaussians is a more deterministic approach. It also points out the inadequacy of the hyper parameters determined in Section \ref{sec:hj_results}: inability to be applied universally. We will discuss this limitation further in Section \ref{sec:limits_hyper}.

\section{Conclusion and Discussion} \label{sec:conclusions}
Our study presents the novel technique of calibrating and exploring the subgrid physics parameter space with machine learning methods. It is computationally expensive to run a simulation and it is prohibitively costly if we intend to fully explore the subgrid models. However, it is a necessary and vital step to bridge the results from simulations and observations. We build on the work of \citet{2020MNRAS.497.5203O} to illustrate the significant savings and increased agreement brought about by the machine learning approach. We summarise our conclusions as follows:

\begin{itemize}
	\item We use a neural density estimator as the basis of the machine that is used to explore the parameter space. It makes use of a mixture of Gaussian distributions to represent the output. Also, it is suitable for an under-determined problem such as calibrating the subgrid physics model for numerical simulations. It is used in parallel with active learning, allowing interactions between data generation and machine. The aim is to focus the exploration in the region that is of interest determined by the machine. See Section \ref{sec:hj_setup} for more details.
	
	\item The machine consists of a number of hyper parameters that require calibration before it can be applied to numerical simulations. As such, we use a projectile motion problem designed with a similar number of inputs and outputs to the simulation for the determination of the hyper parameters. The inputs are the drag coefficient of the thrown object, velocity, and angle of the throw while the outputs are the maximum height and distance of the resulting projectile motion. We tune the number of Gaussian distributions, the number of initial data as a training set, and the number of additional data per iteration of the machine. The final setup consists of 20 Gaussian distributions, five initial training data, and one additional data per iteration. These values are found to minimize the number of throws to obtain a height and distance within 5\% of $2.3\,{\rm m}$ and $0.6\,{\rm m}$ respectively. See Section \ref{sec:hj_results} for more details.  
	
	\item We then apply the machine to match the baryon makeup of a MW-sized halo at $z=0$. As mentioned, the problem is defined with three inputs and two outputs. The inputs are $f_*$, $\epsilon$ and $r\_s$ while the outputs are $\left|\Delta f_{\rm d}\right|$ and $\left|\Delta f_{\rm s}\right|$ (see Section \ref{sec:bk_setup}). We obtain a simulated MW-sized galaxy with $\left|\Delta f_{\rm d}\right| = 0.00642$ and $\left|\Delta f_{\rm s}\right| = 0.0636$ at $z = 0$ using (2.7e-5, 0.85, 1\_1). Compared to the results from \citet{2020MNRAS.497.5203O}, we achieve a better agreement to observations and it is obtained with 22 simulations. This number translates to a factor of three improvements over that in \citet{2020MNRAS.497.5203O}. However, due to the exploratory nature of the machine, it did not continue exploring this region of parameter space, which can potentially yield a better match. See Section \ref{sec:bk_results1} for results and discussion.
	
	\item Since we hope for the machine to continue exploring a particular region of parameter space if deemed able to reproduce observations, we adjusted the strategy used by the machine. Rather than treating each Gaussian peak equally, if the most probable distribution is more than the next by more than 20\%, it is designed to focus on the most probable Gaussian distribution. The recommendation will then be provided from this distribution. The machine identifies a set of parameters in a different region of parameter space (3.0e-5, 0.991, 1\_3) that produces a baryon makeup at $z = 0$ that is in between \citet{2020MNRAS.497.5203O} and that obtained with the previous strategy ($\left|\Delta f_{\rm d}\right| = 0.0428$, $\left|\Delta f_{\rm s}\right| = 0.0708$). However, it takes more simulations than before to achieve this aim. Despite using more computational resources, the machine focuses them in this particular region of parameter space and locates another set of parameters (2.8e-5, 0.996, 1\_3) which gives a better agreement than the first set ($\left|\Delta f_{\rm d}\right| = 0.0328$, $\left|\Delta f_{\rm s}\right| = 0.0541$). This behavior suggests that the modification to the strategy of the machine worked as intended. See Section \ref{sec:bk_results2} for results and discussion.
	
	\item We also demonstrate the versatility and robustness of the machine to be applied to various problems without modifications. It is used to reconcile the results between a grid-based ({\tt Enzo}) and particle-based ({\tt GIZMO}) simulation code as shown in \citet{2021ApJ...917...12S}. For this problem, the inputs that will be varied are the diffusivity and feedback energy per SN of a {\tt GIZMO} simulated isolated disk galaxy. We compare the outputs, which are the $M_{\rm metal}$ and $M_{\rm gas, Z\sim0.01}$ of the {\tt GIZMO} simulated galaxy to that of {\tt Enzo}. A set of parameters is identified with human intervention to produce outputs within 10\% of both simulation codes: $C_{\rm d}\approx 0.0218$ and feedback energy per SN $\approx 1.54\times10^{51}\,{\rm ergs}$. The corresponding outputs of the simulations demonstrate the capability of the machine to produce a better agreement. Also, the results pointed out the sensitivity of the halo metal mass to $C_{\rm d}$ while questioning the relation of $M_{\rm gas, Z\sim0.01}$ to $C_{\rm d}$.
	
	\item In order to obtain a more deterministic solution than human intervention to resolve the oscillatory behaviour encountered in Section \ref{sec:ej_results_human}, we look to modify the values of the hyper parameters. The number of Gaussians is increased from 20 to 100. We showed that this approach effectively suppressed the oscillation in Section \ref{sec:ej_results_machine}. However, this finding hints at the need to determine the optimal combination of hyper parameters tailored to each problem. 
\end{itemize}

\subsection{Limitations of model} \label{sec:limits_model}
Including this work, there have been two attempts at calibrating the \citet{1992ApJ...399L.113C} model of star formation and feedback using $f_*$, $\epsilon$ and $r\_s$ to reproduce the baryon makeup of a MW-sized halo. However, a perfect match still eludes us, suggesting the presence of limitations. It can lie with the parameters used, i.e., the number of degrees of freedom is insufficient. There are more tunable parameters available and it seems they are necessary to improve the match. However, this increment in the number of parameters means a corresponding increment in the complexity, which further motivates the use of the machine learning method developed in this paper. Alternatively, there might be intrinsic limitations at this relatively higher resolution than the intended resolution of the model.

\subsection{Limitations of hyper parameters}  \label{sec:limits_hyper}
While it might not be generic enough to represent the other problems investigated, the hyper parameters determined by the projectile motion problem serve as a beacon to a region of interest in parameter space. However, as we have demonstrated in Section \ref{sec:ej_results_machine}, this combination of hyper parameters is not universal. If provided with substantially more computational resources and time, a tailored set of hyper parameters can be potentially determined for specific problems. Despite this restriction, with the current set of hyper parameters, the method is already showcasing improvements in terms of both the requirement of resources and the accuracy of the simulated properties.

When we fix the hyper parameters setup, we demonstrated the difference from the strategies used to obtain recommendations from the machine. The list explored in this work is definitely not exhaustive and can even be used in combination. The entire process of calibrating the subgrid physics model can be fully automated but it appears that with additional analysis conducted by the user, a better understanding of the model can be obtained. With an improved grasp of the applied model, a better set of parameters can be identified within a shorter amount of time, shifting the focus towards scientific analysis of the results. 

\section*{Acknowledgements}

Ji-hoon Kim acknowledges support by Samsung Science and Technology Foundation under Project Number SSTF-BA1802-04, and by the POSCO Science Fellowship of POSCO TJ Park Foundation. 
His work was also supported by the National Institute of Supercomputing and Network/Korea Institute of Science and Technology Information with supercomputing resources including technical support, grants KSC-2020-CRE-0219 and KSC-2021-CRE-0442.
S.E.H was supported by the project \begin{CJK}{UTF8}{mj}우주거대구조를 이용한 암흑우주 연구\end{CJK} (``Understanding Dark Universe Using Large Scale Structure of the Universe''), funded by the Ministry of Science.
The publicly available {\tt ENZO} and {\tt yt} codes used in this work are the products of collaborative efforts by many independent scientists from numerous institutions around the world. 
Their commitment to open science has helped make this work possible.

\section*{Data Availability}

The data underlying this article will be shared on reasonable request to the corresponding author.




\bibliographystyle{mnras}
\bibliography{references} 



\appendix


\bsp	
\label{lastpage}
\end{document}